# Catalysis in Extreme Field Environments: A Case Study of Strongly Ionized SiO$_2$ Nanoparticle Surfaces

Thomas M. Linker[1,2*], Ritika Dagar[3], Alexandra Feinberg[1], Samuel Sahel-Schackis[1], Ken-ichi Nomura[4], Aiichiro Nakano[4], Fuyuki Shimojo[5], Priya Vashishta[4], Uwe Bergmann[2], Matthias F. Kling[1,6**], Adam M. Summers[1,6***]

1. Stanford PULSE Institute, SLAC National Accelerator Laboratory, Menlo Park, California 94025, US

2. Department of Physics, University of Wisconsin-Madison, Madison, WI, 53706, USA

3. Department of Physics, Ludwig-Maximilians-Universität Munich, D-85748 Garching, Germany

4. Collaboratory for Advanced Computing and Simulations, University of Southern California, Los Angeles, CA 90089-0242, USA

5. Department of Physics, Kumamoto University, Kumamoto 860-8555, Japan.

6. SLAC National Accelerator Laboratory, Menlo Park, CA, 94025, USA

*Corresponding Authors*

\* tlinker@slac.stanford.edu

\*\*mfkling@slac.stanford.edu

\*\*\* asummers@slac.stanford.edu

**Abstract**

High electric fields can significantly alter catalytic environments and the resultant chemical processes. Such fields arise naturally in biological systems but can also be artificially induced through localized nanoscale excitations. Recently, strong-field excitation of dielectric nanoparticles has emerged as an avenue for studying catalysis in highly ionized environments producing extreme electric fields. While the dynamics of laser-driven surface ion emission has been extensively explored, understanding the molecular dynamics leading to fragmentation has remained elusive. Here we employ a multiscale approach performing non-adiabatic quantum molecular dynamics (NAQMD) simulations on hydrogenated silica surfaces in both bare and wetted environments under field conditions mimicking those of an ionized nanoparticle. Our findings indicate that hole localization drives fragmentation dynamics, leading to surface silanol dissociation within 50 femtoseconds and charge transfer-induced water splitting in wetted environments within 150 femtoseconds. Further insight into such ultrafast mechanisms is critical for advancement of catalysis on the surface of charged nanosystems.

**Main**

The influence of electric fields on molecular and electronic structures has been a subject of scientific inquiry since the pioneering work of Stark, who over a century ago documented field-induced shifts in electronic energy levels[1]. Recent research has expanded our understanding of how electric fields affect chemical bonding with significant insights into their role in driving localized bond breaking only emerging in the past few decades[2–4]. Notably, extreme electric fields have been known to emerge from local gradients/dipolar structures in biological systems, with magnitudes up to ~1V/Å[5]. However, the role of these fields in catalysis reactions has only relatively recently come to light[6–9]. Rare reaction channels, such as water autoprotolysis ($2H_2O \rightarrow H_3O^+ + OH^-$), has been attributed to spontaneous dipolar arrangements that lead to the generation of large local electric fields (reaching magnitudes of up to 3V/Å), underscoring the potential of strong local electric fields to drive unique chemical dynamics[10–12].

Building upon these principles, recent advancements in catalysis and efforts to drive increases in catalytic rates have sought to harness strong local electric fields, generated through both direct methods such as scanning tunneling microscopes (STM) as well as indirect methods through the creation of chemical environments with strong local electric fields[13–15]. Engineering of catalytic processes is pivotal in various industrial and environmental applications, including energy conversion, pollution mitigation, the synthesis of fine chemicals, and even cancer treatments[16–21]

Another direct method to generate strong local electric fields is light manipulation of nanoparticles, which can generate ionized particle surfaces with correspondingly strong electric fields. Historically, photo catalysis at nanoscale has focused on localized surface plasmon resonance (LSPR) in metal nanoparticles, and has been predominantly associated with noble metal nanoparticles, such as gold and silver, which exhibit strong LSPR in the visible spectrum[18]. However, the inherent energy losses and non-negligible heat generation associated with these materials under laser excitation have prompted the exploration of alternative platforms. Dielectric nanoparticles, such as silica, have been proposed as an energy-efficient alternative, capable of inducing catalytic reactions through mechanisms like strong field ionization without the drawback of excessive thermal effects[22]. Recent advancements have highlighted the induction of catalytic reactions on silica nanoparticles through strong field ionization, leading to the generation of free charges that instigate a cascade of chemical dynamics[23–29]. Notably, the formation of trihydrogen cations ($H_3^+$) from the dissociation of molecular adsorbates on silica surfaces has been demonstrated, unveiling a novel inorganic pathway for $H_3^+$ formation[25]. $H_3^+$ serves as a hallmark for water splitting and as a precursor for forming complex organic compounds[30]. Understanding the pathways that give rise to trihydrogen cations and similar reaction channels can not only give insight into the use of strong field excitation as a method for photocatalysis, it can also elucidate the general chemistry in extreme field conditions. The strong-field excitation not only creates free charge carriers that directly partake in molecular fragmentation, but also extreme electric field conditions on the surface of the nanoparticle. As up to tens of thousands electrons can be ejected from nanoparticles upon strong field ionization, local fields reaching ~1 V/Å can be achieved at the particle surface[23–29]. Since silica is largely non-catalytic prior to ultrafast laser excitation, the ability to rapidly induce strong, and highly localized electric fields, allows for controllable and time-resolved studies of the mechanisms for catalysis in such strong fields with free charge carriers.

Despite the recent progress, understanding the precise role that surface charges and local fields play on the dynamics of chemical bonding and the ultrafast timescales on which these interactions occur in nanoscale systems has remained elusive. Here we adopt a multiscale simulation approach to investigate the molecular fragmentation dynamics resulting from strong field ionization of wetted silica nano surfaces. The multiscale approach is needed as first principles molecular simulations based on density functional theory (DFT) can at most simulate a few thousand atoms, while typical nanoparticles used in experiments exceed 25 nm and thus contain millions of atoms or more. While the simulations presented here are performed for ionized silica nanoparticles, the mechanisms and timescales uncovered for molecular fragmentation dynamics are anticipated to be relevant to many avenues of catalysis in strong field environments with free charge carriers.

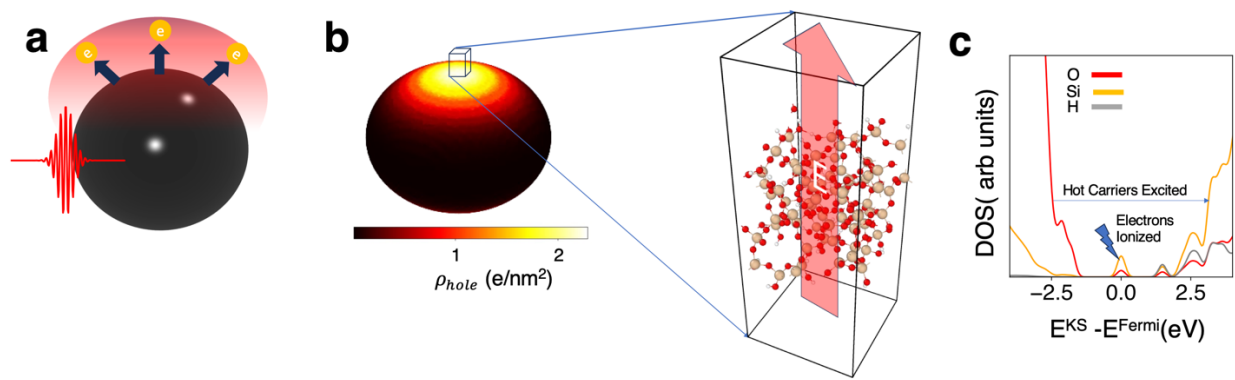

**Figure 1.** (a) Diagram of strong field induced electronic ionization. Due to near field enhancement of the ionizing laser, the density ionized electrons is much higher at the poles. (b) Resultant non uniform charge density on the nano sphere from ionization from the optical pump pulse. This ionization density creates a strong electric field on the surface of the nano sphere. We use the range of electric fields strengths on the surface of the sphere as boundary conditions for NAQMD simulations on a model amorphous silica surface slab. Si atoms are colored tan, oxygen red, and hydrogen white. (c) Element-projected Kohn-Sham (KS) density of states of the model amorphous silica slab with relevant excitations indicated.

In the first step of this multiscale approach, we calculate the electric field formed on the ionized silica surface. We first compute the hole/charge distribution generated on the surface of a nano-sphere through strong field ionization from a typical ultrashort 800 nm, 5 fs laser pulse.

Further information on these strong-field ionization simulations can be found in the SI. These conditions are very similar to those reported in previous experimental studies on the strong-field ionization of nanoparticles[23–29]. Due to near-field enhancement effects from the Mie-scattering of the laser pulse with the nano-sphere and the nonlinear nature of strong field ionization, the distribution is non uniform with high concentration of ions at the poles of the sphere. A diagram of this process is illustrated in figure 1a. While the initial conditions were chosen to resemble previous experimental work, the results discussed below apply more broadly to other charge distributions on nanosurfaces.

Figure 1b shows the positive (hole) charge density formed on the surface of a silica sphere with a 50 nm radius within the 5 fs duration for a laser pulse with 5 x $10^{13}$ W/cm$^2$ intensity. More details of how this density is computed is described in the SI (see figures S1 and S2). The ionization rate, and thus the surface charge density is maximized at the poles and falls off with polar angle. Within this 5 fs window, the atoms on the surface have experienced little motion and are now exposed to an extremely strong electric field resulting from the positive charge distribution on the particle surface. This is the starting point for our molecular simulations, where we use the range of electric field strengths on the nanosphere surface as field boundary conditions for a small ~1 nm$^2$ amorphous hydrogen terminated silica slab. The amorphous slab is visualized in figure 1b. We find the largest magnitude of the field at any point on this surface is on the order of ~2 V/Å and the lowest on the order of 0.2 V/Å (see supplemental figure S2). The local fields are consistent with what is anticipated from electron emission measurements.[26]

Thus, to understand the molecular fragmentation dynamics, NAQMD simulations[31,32] in the presence of a static electric field are performed on a small ~1 nm$^2$ hydrogen terminated amorphous silica surface. The application of a strong DC field along an amorphous silica surface to model the effect of the fields induced by the global hole density on the nanosphere is similar to globally informed Hartree potential frameworks in divide and conquer density functional theory[33]. NAQMD simulations are based on time-dependent DFT (TD-DFT) within the surface hopping approach[34]. Calculations are performed in a plane-wave basis with the projected augmented wavevector (PAW) method[35]. The Perdew-Burke-Ernzerhof (PBE) version of generalized gradient approximation (GGA)[36] was used and van der Waals corrections were employed utilizing the DFT-D scheme[37]. (Further details are given in the SI). We consider the ionization of electrons as well as the high energy electron-hole pairs (i.e., non-ionizing excitations) which can occur due to local Coulomb trapping[26]. Partial densities of states projected onto elements (O, Si, and H) are also plotted for the amorphous silica slab in figure 1c with relevant excitations illustrated. In the amorphous slab, a gap state is formed at the Fermi-level. This is expected to occur in wetted silica slabs with dangling bonds[38,39].

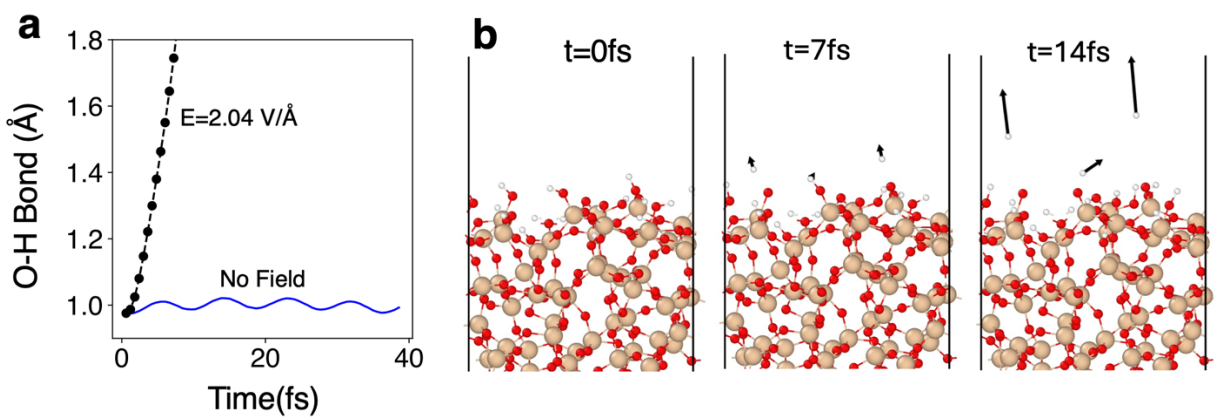

**Figure 2.** (a) O-H bond dynamics for surface selected Silaonal group in presence of large electric field where ultrafast dissociation was seen. (b) Snap shots of MD trajectory under strong electric field with arrows drawn for protons quickly escaping the surface.

We first examined the effects of the largest field strengths on the surface of the nanosphere. When the electric field is on the order of ~2 V/Å the dissociation of O-H bonds at the surface is nearly instantaneous (<~10 fs) regardless of whether any local charge carriers are present. Figure 2a illustrates the O-H bond dynamics at the hydrogen terminated silica surface under a 2 V/Å field for a given O-H bond on the surface compared to the ground state dynamics, for which ultrafast dissociation can be seen. Snapshots of the MD trajectory are shown in Figure 2b where within 24 fs, 3 protons are ejected. For such strongly ionized surfaces and their corresponding strong electric fields, the dynamics are essentially limited by the O-H stretching frequency.

At a weaker field strength close to 1 V/Å, we found the dynamics become more subtle and are highly dependent on the spatial distribution of the excitation as well as the local hole density. In this regime, we first considered the local field dependence for a region with a high distribution of excited carriers with the valence band being doubly excited to a high lying electronic state and the gap state empty (three holes and two excited electrons, net ionization state of +1). For these simulations we found the excited electrons were highly delocalized and did not contribute to the dissociation dynamics beyond thermalization from carrier-phonon scattering, and all dissociation dynamics was driven by the holes. If the valence band maximum (VBM) was found to be initially localized to the surface of the slab, the excited hole wave-function could quickly localize to an individual oxygen atom resulting in dissociation of the Si-O bond and formation of an $OH^+$ ion within 30 fs as illustrated in figure 3a. The localization dynamics of the VBM wave function is plotted in figure 3b, showing the maximum Mulliken population of any given atom for VBM wavefunction as function of time. Here, the value of unity indicates that the VBM wavefunction is entirely localized to one atom. Figure 3b illustrates that localization of the VBM wavefunction is crucial to dissociation. Such localization has similarly been proposed as the rate limiting step in strong-field dissociation of water [40]. It was found that gap state hole was primarily localized to bare O and Si atoms and did not contribute to any dissociation (see figure S3 in SI), indicating that such states while contributing to the overall global charge state of the ionized nano-sphere can locally suppress dissociation by trapping carriers that would otherwise contribute to bond-breaking.

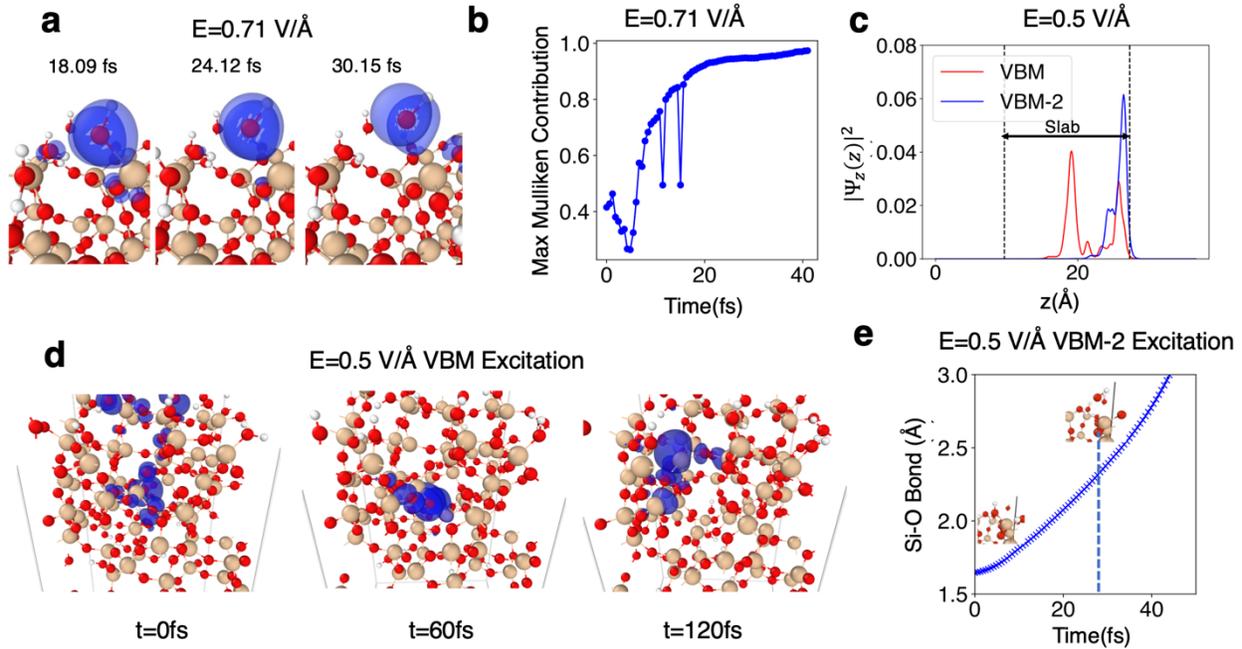

**Figure 3.** (a) Dissociation dynamics at 0.71 V/Å. Localization of hole (blue iso-surface) to single Oxygen atom on Silanol group results in breaking of Si-O bond and formation of $OH^+$. (b) Localization dynamics of the hole wave function as a function of time quantified by the maximum Mulliken population of any given atom to the hole wave function. A value of one indicates complete localization. (c) VBM and VBM-2 wavefunctions for an electric field of 0.5 V/Å. VBM-2 shows primarily surface localization while the VBM is hybridized between the surface and the bulk. (d) Hybridized surface-bulk VBM excitation breaks SiO4 tetrahedra resulting in fast diffusion of the $SiO3^+$ complex within the bulk of the slab. (e) Surface VBM-2 excitation results in rapid breaking of Si-OH bond.

In fields below 0.71 V/Å we found the VBM was not initially localized to surface atoms but primarily localized to the bulk of the slab, with the z dependence of the VBM probability distribution plotted in figure 3c for an electric field of 0.5 V/Å. We found that under these conditions the dynamics would not result in fast bond dissociation (<100 fs). Here, the holes would primarily localize to an $SiO_4$ tetrahedra in the center of the slab resulting in fast diffusion of these atoms. Figure 3d shows the initial atomic configuration with the VBM wavefunction hybridized between the surface and the bulk atoms. The excited VBM hole wave-function then eventually localizes to $SiO_4$ tetrahedra within the center slab, resulting in the tetrahedra breaking as illustrated in figure 3d. We observed similar dynamics when exciting from a bulk like valence state with a field strength of 0.71 V/Å.

Since the difference between the VBM, VBM-1 and VBM-2 energies was less than 100 meV, we also considered excitations from these states if the initial wave-functions were more localized to surface atoms. The *z* dependence of the VBM-2 wave-function for a field of 0.5 V/Å is illustrated in figure 3c, which is localized to the surface of the slab. We found down to a field of 0.5 V/Å that when excited from surface localized state, Si-OH disassociation would occur within 30 fs, with dynamics plotted in figure 3e. The dissociation for fields of 1.02 V/Å, 0.71 V/Å and 0.5 V/Å was found to be on the order of 30 fs for a surface like excitation. Below 0.5 V/Å field we did not observe any dissociation within 240 fs of simulation.

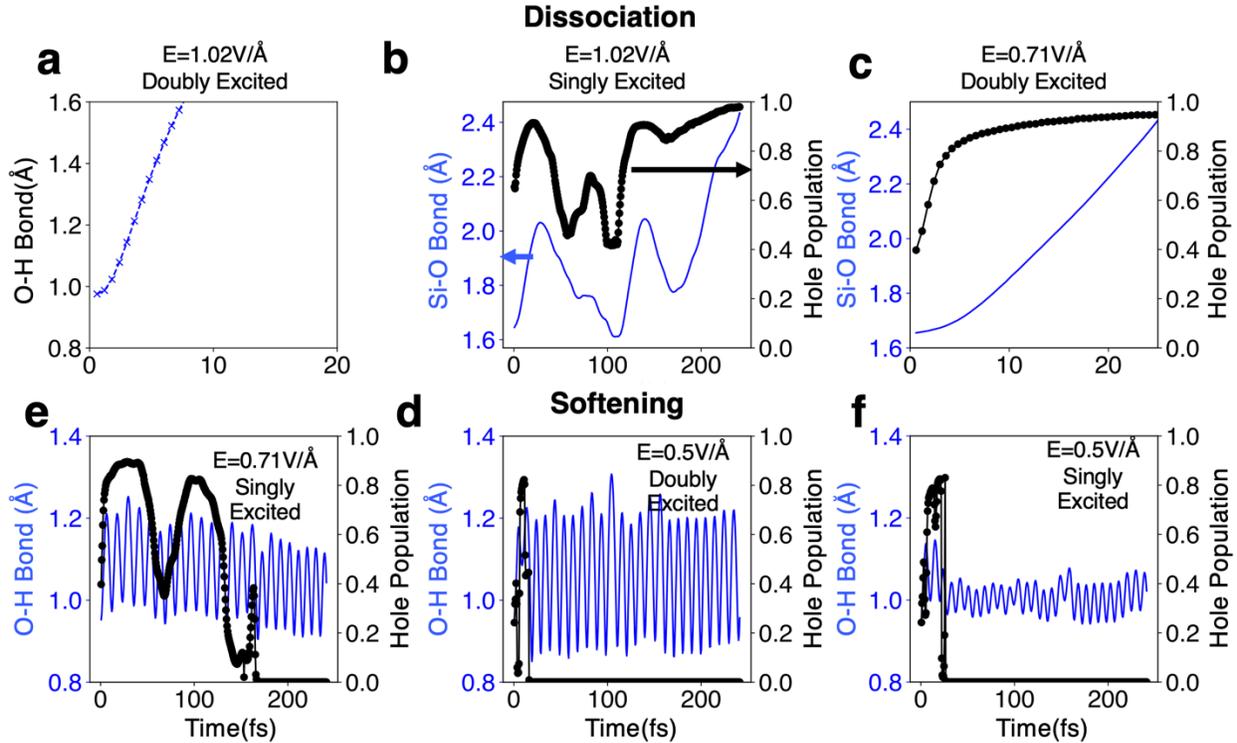

**Figure 4.** (a) Ultrafast dissociation of O-H bond in doubly excited surface valence state with filled gap state and electric field of 1.02 V/Å. (b) Same conditions as (a) but singly excited valence state. This results in a much longer dissociation process of a surface Silanol group bond resulting in OH$^+$. Long dissociation process is a result of long hole localization time. (c) Same conditions as (a) but lower field of 0.71 V/Å resulting in ultrafast dissociation of Si-OH bond resulting in OH$^+$. Similar to the dynamics observed with gap state unoccupied as described in figure 3. (d) Same conditions as (c) but singly excited state resulting in softening of O-H bond with long period of hole localization. O-H bond dynamics slowly relax after the hole localizes to the bulk. (e) Same conditions as (a) but lower field of 0.5 V/Å. Brief hole localization on a surface O atom results in O-H bond softening. The O-H bond remains softened well after the hole wave-function localizes to the surface. (f) Same conditions as (e) but singly excited hole wave-functions. Brief hole localization slightly disturbs O-H bond dynamics before relaxing back to near ground state vibrational stretching.

Next we examined the effect of decreasing the carrier concentration at the examined fields, by first filling the gap state, and then decreasing the double valence excitation to a single excitation. The excitations were taken from wave functions near the valence edge that were localized to the surface. For a field strength of 1.02 V/Å, we found that ultrafast dissociation still occurred with a double valence excitation and the gap state filled, with fast proton ejection observed within 10 fs as illustrated in figure 4a. For a single excitation we found a much longer dissociation process, with Si-OH dissociation occurring within 220 fs as illustrated in figure 4b.

The long dissociation process was a result of long localization time of the hole wave-function to a surface oxygen atom. During this time, the periods of partial hole localization to the surface oxygen atom involved in the dissociation process led to softening of the Si-O bonds as well as the O-H bond in the involved in the dissociating silanol group (see figure S4 in SI).

When reducing the field to 0.71 V/Å and doubly exciting a surface valence wave-function, we saw an ultrafast dissociation process within ~20 fs as illustrated in figure 4c. This is similar to the case when doubly exciting with an unoccupied gap state. In contrast when the same valence state was singly excited, we did not observe any dissociation. We found the hole wave-function would begin with its maximum Mulliken population on a surface oxygen atom before eventually localizing to the bulk in ~100 fs, which is illustrated in figure 4d. This brief period on the surface oxygen atom resulted in a long period of softening and slow relaxation of the corresponding O-H stretching vibration due to the deposited kinetic energy from the localization period as shown in figure 4d. At a field strength of 0.5V/Å, we saw similar softening dynamics without dissociation as result of brief hole surface localization when doubly exciting a surface valence electron, as illustrated in figure 4e. In contrast when the gap state was unoccupied the same double excitation led to ultrafast dissociation. This indicates that while the gap state we have found to localize to bare O and Si atoms at the surface and not the silanol groups, it can still contribute to silanol dissociation through its interaction with other holes. When only singly excited, a similar brief hole localization only led to brief elongation of the O-H bond before relaxing to near ground state stretching dynamics as illustrated in figure 4f (see supplemental figure S4 for example of ground state stretching dynamics).

Overall, these results indicate for a wetted silica surface a threshold behavior for ultrafast dissociation occurs in terms of both local charge density and the global charge state. Such threshold-like behavior has similarly been observed in charge localization and resulting bond dissociation in NAQMD investigations of dielectric breakdown[40]. We note these simulations cannot capture large spatiotemporal dynamics and effects such as electron-hole recombination and charge diffusion are also anticipated to play a role. However, the results clearly show the necessity for surface charge localization to initiate fragmentation on bare silanol groups. In addition, in weaker excitation regimes brief periods of surface localization on surface oxygen atoms result in substantial softening of affected O-H bonds, which can lead to fragmentation if they interact with other surface molecules.

Next we examined the dynamics in a wetted environment with free water molecules present on the surface. Figure 5a shows the amorphous silica surface with 27 water molecules deposited on top. As we found that excitations within the bulk of the bare slab would not result in fast dissociation processes, we considered both excitations from the bulk of the slab and those hybridized with the surface water molecules, with their spatial profiles plotted in figure 5b. In both cases we considered an excitation condition ($E = 0.5$ V/Å, doubly excited valence excitation, and empty gap state) where in the bare slab case the spatial dependence of the excited hole wave function determined if molecular fragmentation occurred. The gap state was again not found to contribute to dissociation. We found that in both excitation cases the holes transfer from silica to free water molecules resulting in water splitting events. For the hybrid like excitation, we found that the two holes would localize on a single water molecule resulting in a $H_2O^{2+} + H_2O \rightarrow OH^+ + H_3O^+$ reaction. The bottom part of figure 5c shows the molecular charges computed via Mulliken analysis for the resulting $OH^+$ and $H_3O^+$ fragments as a function of time with 4 labeled points. The top part of figure 5c function show trajectory snapshots for each of the labeled points illustrating initial uncharged molecules (i), two-hole localization to a single water molecule (ii), charge transfer and dissociation (iii), and ejection of the $OH^+$ from the surface (iv). We plot similar dynamics for the bulk-like excitation in figure 5d with the Mulliken charge dynamics for a water molecule that undergoes hole localization and splitting with 4 relevant labeled points in time. The top part of figure 5d shows snapshots for the labeled points in time illustrating the initial molecule (i), water splitting reaction with $H_3O^+$ and OH formation (ii), recombination with the OH radical (iii), and the $H_2O^+$ molecule leaving the surface(iv). During the simulation another water splitting reaction occurred prior to complete ejection of the $H_2O^+$ molecule (which effectively ends the simulation), resulting in $H_3O^+$ and $H_2O^+$ molecules. The final configuration is plotted in supplemental figure S5. We see in both excitation cases that the silica surface could act as a charge donor resulting in water splitting reactions within less than 150 fs.

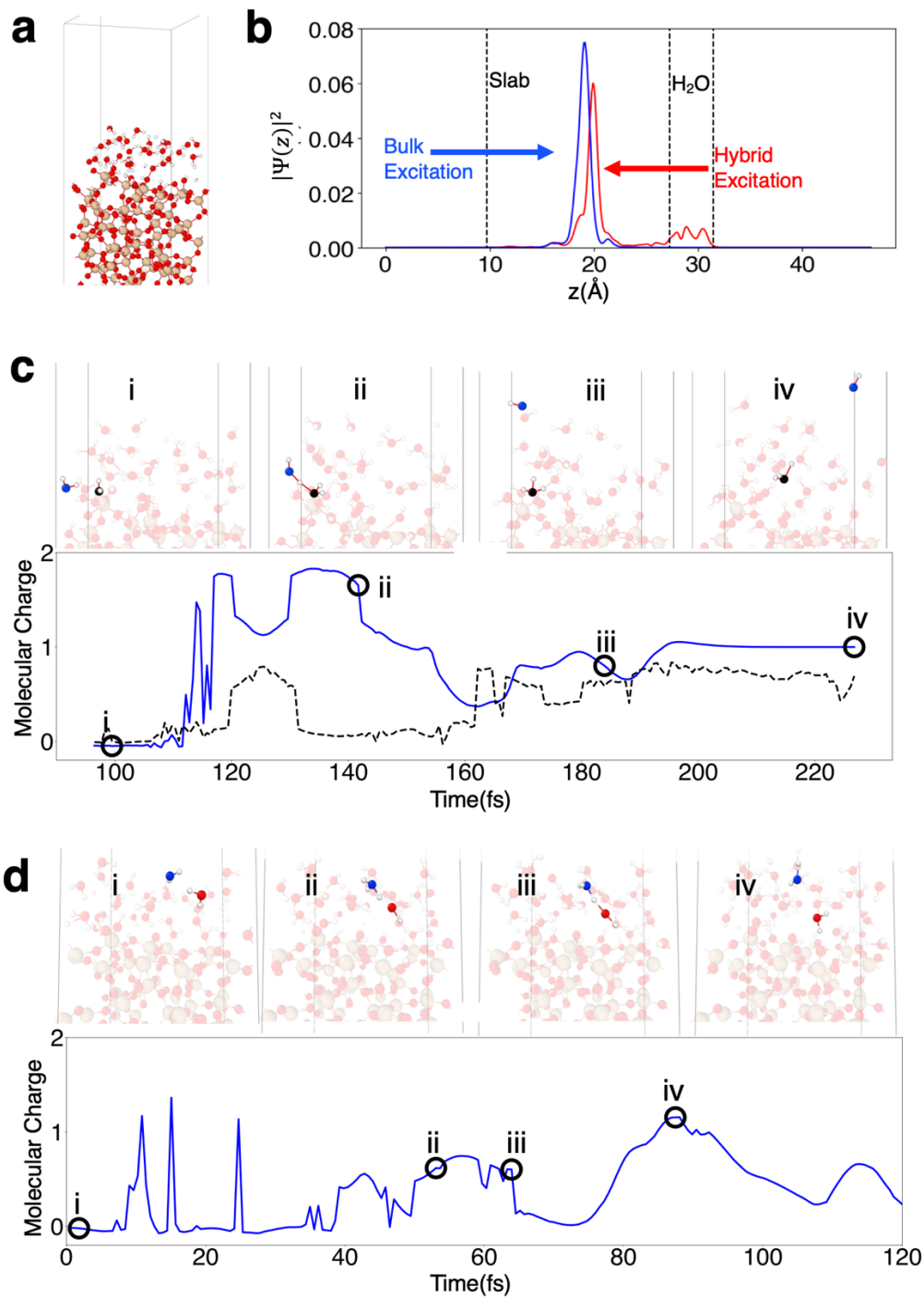

**Figure 5.** (a) Simulated silica water interface. Periodic boundary box shown with white lines(b) Initial hole wavefunctions excited spatial distributions. (c) Molecular snap shots and local charges computed via Mulliken analysis for two water molecules that undergo water splitting reaction for hybrid excitation. Labeled points on the plot correspond to snapshots of the molecular dynamics trajectory plotted above during formation and splitting of $H_2O^{2+}$ to form $OH^+$ and $H_3O^+$ fragments. (d) Molecular snap shots and local charges computed via Mulliken analysis for a water molecule that undergoes hole localization, splitting and recombination during the simulation for the bulk-like excitation. Labeled points correspond to snapshots from the MD simulation plotted above illustrating the dissociation, recombination, and ejection of the water molecule from surface.

    The results illustrate that charge transfer to free molecular species plays a significant role in catalytic activity in ionized dielectric nanoparticles. Excitations that would have led to longer charge diffusion processes in bare silica resulted in fragmentation of the surface molecules. Due to the strong electric fields generated at the surface of the ionized nanosphere, the surface molecules' electronic states are closer to the VBM for the total system. Thus, any high energy holes excited from the silica slab will want to relax to these states.  In this regard the highly ionized dielectric nanoparticle behaves analogously to a metal and can donate free hole carriers to molecular adsorbates to induce a catalytic reaction.

In conclusion, our simulations elucidate the role of charge transfer and localization in strong field catalysis of SiO$_2$ nanoparticles, with charge localization appearing to be the critical limiting step for initializing catalytic activity. In the bare wetted particle, we find primarily a threshold hold like dependence on catalytic activity in terms of the global ionization field and local charge density. In strong field/ excitation regimes we see extremely fast fragmentation dynamics with sub 50 fs silanol dissociation for bare surfaces and ~100 fs charge transfer induced water splitting when free molecules were located at the surface. While in lower excitation regimes we did not observe fragmentation dynamics, we did see significant bond softening which could lead to longer time reaction channels. Further understanding of such reaction pathways will be critical for development of strong field ionization as a tool for catalysis as well as understanding chemistry in strong field environments. The on-demand metallization by strong field ionization of dielectric nanoparticles also offers the ability to time resolve these processes experimentally. Such information is difficult to obtain in metal nanoparticles that are catalytic prior to laser excitation.

## Acknowledgments

The work at SLAC and the University of Wisconsin was supported by the U.S. Department of Energy, Office of Science, Basic Energy Sciences, Chemical Sciences, Geosciences and Biosciences Division. R.D. and M.F.K. acknowledge support from the German Research Foundation (DFG) for simulations of the strong-field ionization on nanoparticles. The NAQMD simulations were performed at the Stanford Sherlock Cluster and the University of Southern California Advanced Research Computing Center.

## Author Contributions :

T.M.L, R.D, A.M.S, and M.F.K designed the multi-scale simulation framework. T.M.L, A.N., P.V., F.S., designed the NAQMD simulations supported by discussions with R.D. and A.M.S. T.M.L performed the NAQMD simulations. R.D and T.M.L performed the strong field ionization simulations. T.M.L and A.M.S wrote the first draft of the manuscript. All authors contributed to the discussion of the results, and the final version of the manuscript.

**Supplementary Information**

***Theory for Strong Field Ionization***

We utilized Ammosov, Delone, and Krainov (ADK) theory to estimate the charge density generated on a Silica nanosphere for a typical laser pulse used in ultrafast experiments (800 nm 5fs, 5 x $10^{13}$ W/cm$^2$)[1]. We integrate ADK rate over the pulse duration to determine the probability density of ionization for any given surface area element on the sphere. We compute the total ionized charge density as density of "SiO2" in crystalline silica times the ionization probability[1]. Fig. S1a shows the charge distribution at the surface as a function of the polar angle θ for a sphere of radius 50 nm. To obtain the electric potential Ψ(r,θ) we solve poisson's equation via Legendre polynomial expansion (azimuthal symmetry is assumed) whose convergence with respect to polynomial order $l$ is shown in figure S1b. The corresponding electric field is plotted in figure S2.

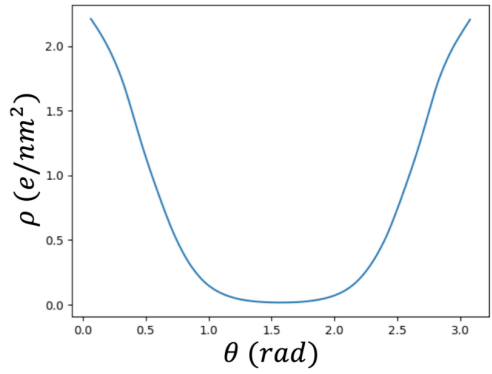 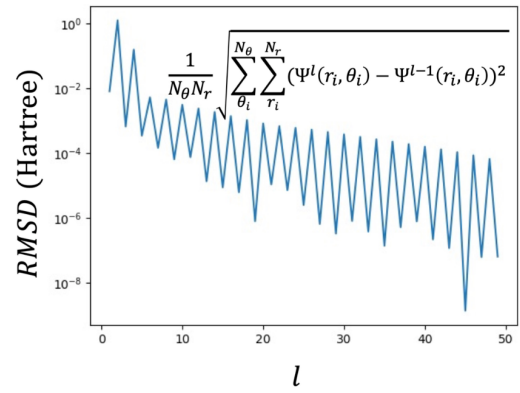

Figure S1. (a) ionization distribution solved from ADK theory. (b) Convergence of the electric potential with respect to Legendre polynomial expansion of order $l$.

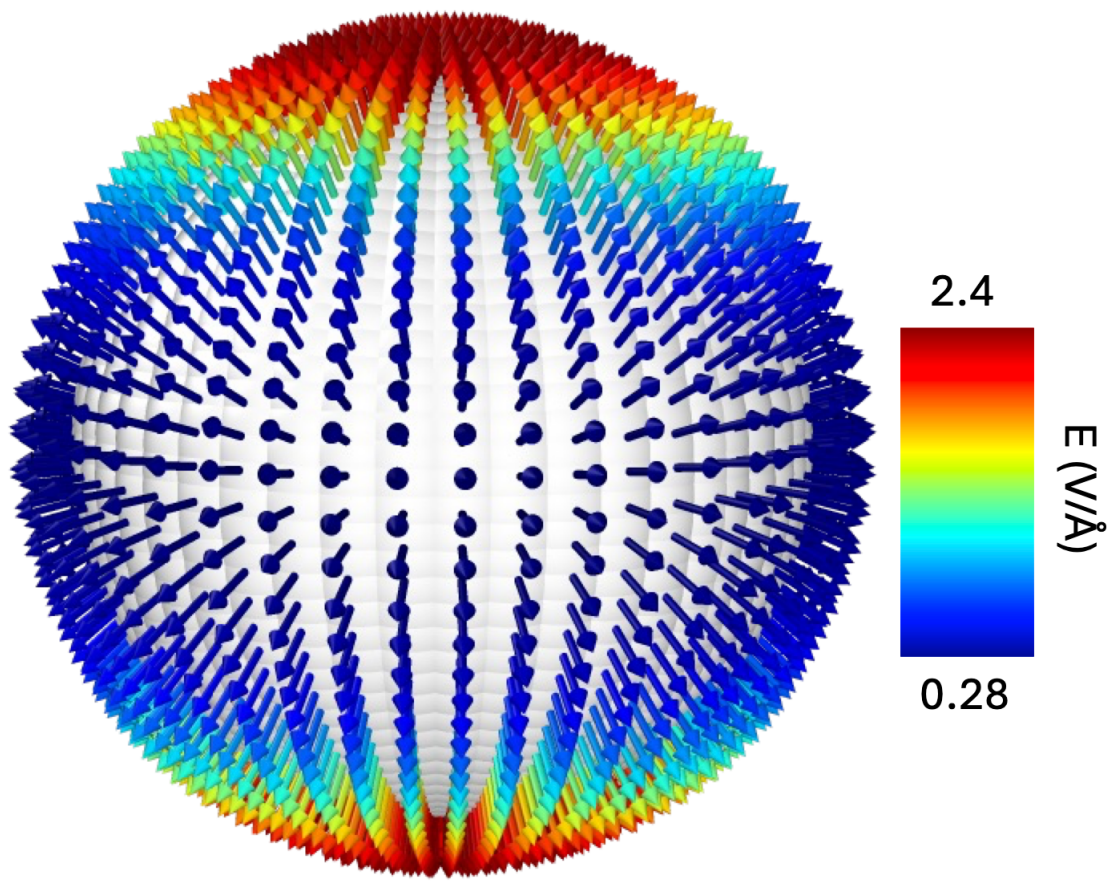

Figure S2. Electric field distribution on the sphere.

**NAQMD Simulations**

NAQMD simulations were performed on a 14.32Å cubic length amorphous Silica structure. Amorphous silica structure was prepared utilizing a classical MD heat and quench simulations with a Vashishta style pair potential[2,3]. Along one of the cubic axes, we created a hydrogen terminated surface and added additional vacuum to prevent image interactions. This structure was relaxed via the conjugate gradient method for 200 steps with density functional theory(DFT) and then thermalized for 500 steps at 300K in DFT-MD prior NAQMD simulations. NAQMD simulations were then performed in the microcanonical ensemble. A time step of 25 a.u (0.603fs) was used in all MD simulations. The electric fields were applied using a sawtooth potential. DFT-MD/NAQMD simulations are ab-initio molecular dynamics approaches that integrate the trajectories of all atoms by computing their intermolecular forces from first principles in the framework of density DFT. NAQMD allows for dynamics of excited carriers and corresponding excited state forces to be modeled within the framework of time dependent DFT[4,5]. Excited state transitions are modeled within the fewest switch surface hopping method[6] To approximate the exchange−correlation functional in our DFT approach we used the Perdew-Burke-Ernzerhof (PBE) version of generalized gradient approximation (GGA)[7]. Van der Waals corrections were employed utilizing the DFT-D scheme[8]. The projected augmented wavevector (PAW) method was used to calculate electronic states within a plane wave basis set[9]. Projector functions were generated for the Si 3s and 3p states, O 2s and 2p states, and the 1s state for hydrogen. A planewave cutoff energy of 35 Ry was used. The NAQMD algorithm was implemented in QXMD quantum molecular dynamics simulation code[5]. For further details of the NAQMD algorithm, see supplementary refs 4,5.

**Gap state**

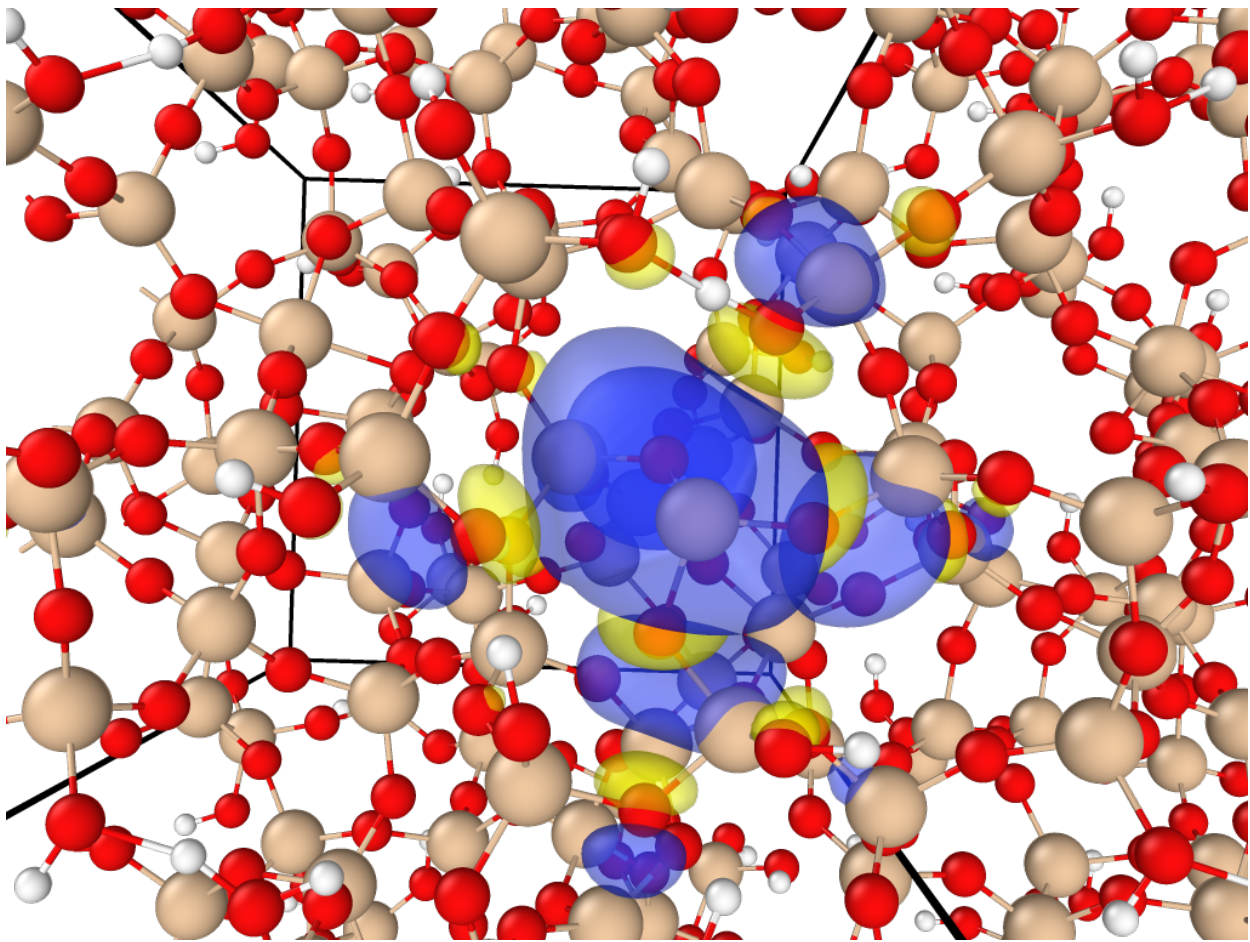

Figure S3. Top view of gap state wavefunction isosurface which is primarily localized to exposed Si surface atom. We found this state to remain localized and not contribute to dissociation.

*Bond Softening of O-H bond during Si-OH dislocation,*

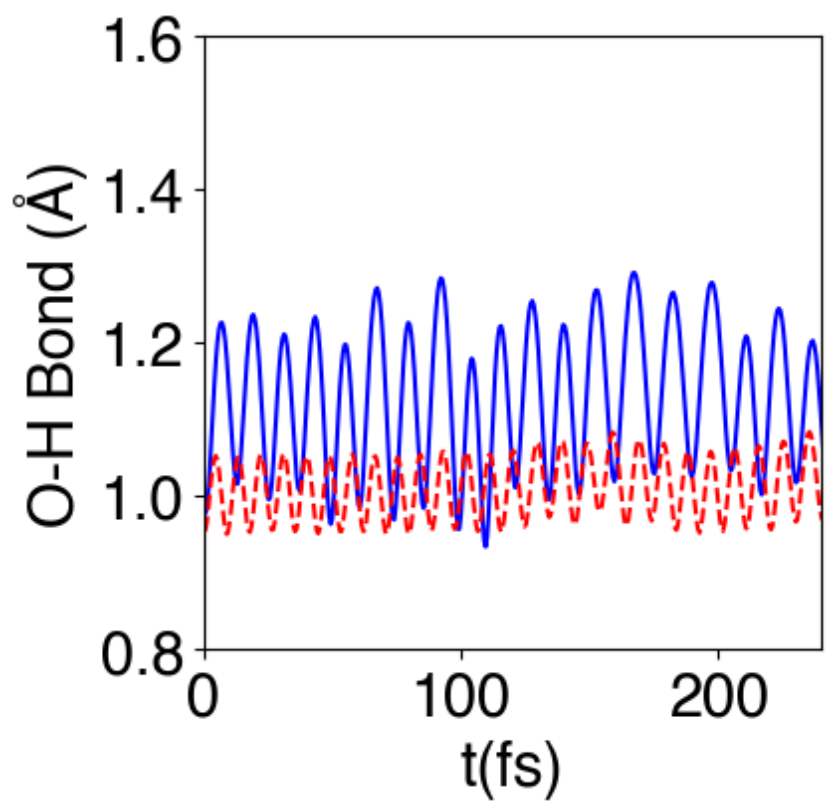

Figure S4. O-H bond dynamics during dissociation of Si-OH bond plotted with straight blue line. For comparison ground state dynamics are plotted with a dashed red line. During the breaking of the Si-O bond the O-H bond is strongly softened.

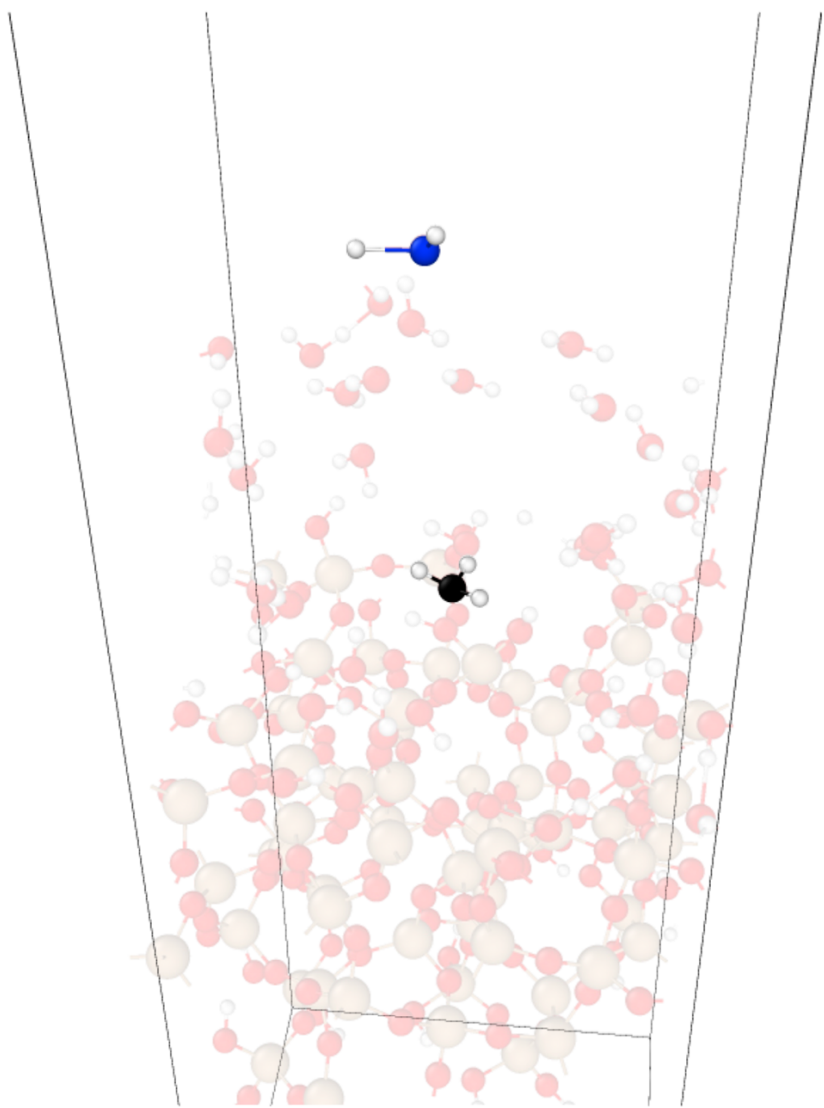

Figure S5. Formation of H3O$^+$ and H$_2$O+ at the end of the simulation of the wetted silica surface using a bulk like excitation.

**Supplemental Information References**